\documentstyle[12pt]{article}

\newcommand{\uqg}{\mbox{$U_q${\bf g}}}
\newcommand{\fun}{\mbox{Fun({\bf G}$_q$)}}
\newcommand{\tr}{\triangleright}
\newcommand{\cross}{\mbox{$\times\!\rule{0.3pt}{1.1ex}\,$}}
\newcommand{\A}{\mbox{$\cal A$}}
\newcommand{\U}{\mbox{$\cal U$}}
\newcommand{\DA}{\Delta_{\cal A}}
\newcommand{\AD}{{}_{\cal A}\Delta}
\newcommand{\I}{\mbox{\boldmath $i$}}
\def\Li{\hbox{\large\it \pounds}}
\newcommand{\dg}{\mbox{\bf d}}
\newcommand{\om}{\mbox{$\omega$}}
\newcommand{\al}{\mbox{$\alpha$}}
\newcommand{\z}{\hspace*{9mm}}
\newcommand{\x}{\hspace{3mm}}
\newcommand{\ad}{\stackrel{\mbox{\scriptsize ad}}{\triangleright}}
\begin{document}
\begin{titlepage}
\begin{center}
June 3, 1993     \hfill    NSF-ITP-93-75\\
                 \hfill    LBL-34215 \\
                 \hfill    UCB-PTH-93/20 \\

\vskip .25in

{\large \bf Cartan Calculus for Hopf Algebras and Quantum Groups}
\footnote{This work was supported in part by the Director, Office of
Energy Research, Office of High Energy and Nuclear Physics, Division of
High Energy Physics of the U.S. Department of Energy under Contract
DE-AC03-76SF00098 and in part by the National Science Foundation under
grants PHY90-21139 and PHY89-04035.}

\vskip .25in

Peter Schupp${}^{1,2}$, Paul Watts${}^2$, and Bruno Zumino${}^{1,2}$ \\[.25in]

{\em  ${}^1$ Institute for Theoretical Physics,
      University of California\\
      Santa Barbara, California 93106-4030\\

      ${}^2$ Department of Physics,
      University of California,	and\\
      Theoretical Physics Group,
      Physics Division\\
      Lawrence Berkeley Laboratory,
      1 Cyclotron Road\\
      Berkeley, California 94720}
\end{center}

\vskip .25in

\begin{abstract}
A generalization of the differential geometry of forms and vector
fields to the case of quantum Lie algebras is given. In an abstract
formulation that incorporates many existing examples of
differential geometry on quantum groups, we combine an exterior
derivative, inner derivations, Lie derivatives, forms and functions
all into one big algebra. In particular we find a generalized Cartan
identity that holds on the whole quantum universal enveloping algebra
of the left-invariant vector fields and implicit commutation relations
for a left-invariant basis of 1-forms.
\end{abstract}

\end{titlepage}

\newpage
\renewcommand{\thepage}{\arabic{page}}
\setcounter{page}{1}

\section{Introduction}

The question of how to endow a quantum group with a differential
geometry has been studied extensively \cite{W2,RTF,SWZ,B,CC,CW}.
Most of these approaches,
however, are rather specific: many papers deal with the subject by
considering the quantum group in question as defined by its R-matrix,
and others limit themselves to particular cases.  Here we
shall attempt a more abstract formulation which depends
primarily on the underlying Hopf algebraic structure; this will
therefore  be a generalization of many previously obtained
results.

The approach we take starts with a Hopf algebra, which we identify
with the (quantum) universal enveloping algebra (UEA) of some Lie algebra,
taking its dual Hopf algebra to be the functions on the corresponding
quantum group.  We then construct a larger (non-Hopf) algebra which
contains these as subalgebras combined by the ``cross
product'' \cite{Md1,SW,SWZ3}.  The differential geometry is then
introduced by
including in the algebra an exterior derivative, Lie derivatives, and inner
derivations.  (Much of the previous work in this area has emphasized
the {\em actions} of these operators on functions and forms rather than
treating them as elements in an extended algebra containing the cross
product algebra and giving the appropriate commutation relations).
Our approach is constructive in nature; this implies that
not only must we treat each given Hopf algebra on a case-by-case
basis, but that questions concerning uniqueness and even existence
arise.  These problems will be addressed in section 3.

\section{Differential Geometry on Hopf Algebras}

\subsection{Hopf Algebras}

A Hopf algebra \A\ \cite{Md1,A,Sw,Df} is an
associative unital algebra $(\A,\cdot,+,k)$ over a field $k$,
equipped with a coproduct $\Delta:\A \rightarrow \A \otimes \A$, an
antipode $S:\A \rightarrow \A$, and a counit $\epsilon:\A \rightarrow
k$, satisfying $(\Delta \otimes i\!d)\Delta(a) = (i\!d \otimes
\Delta)\Delta(a)$, $(\epsilon \otimes i\!d)\Delta(a) = (i\!d
\otimes \epsilon)\Delta(a) = a$, and
$\cdot(S \otimes i\!d)\Delta(a) = \cdot(i\!d \otimes S)\Delta(a) = 1
\epsilon(a)$, for all $a \in \A$. These axioms are dual to the axioms
of an algebra.  There are also a number of consistency conditions
between the algebra and the coalgebra structure: $\Delta(a b) =
\Delta(a) \Delta(b)$, $\epsilon(a b) = \epsilon(a) \epsilon(b)$, $S(a
b) = S(b) S(a)$, $\Delta(S(a)) = \tau (S \otimes S)\Delta(a)$, where
$\tau(a \otimes b) \equiv b \otimes a$ is the twist map,
$\epsilon(S(a)) = \epsilon(a)$, and $\Delta(1) = 1 \otimes 1$, $S(1) =
1$, $\epsilon(1) = 1_{k}$, for all $a,b \in \A$. We will often use
Sweedler's \cite{Sw} notation for the coproduct:
\begin{equation}
\begin{array}{l}
\Delta(a) \equiv a_{(1)} \otimes a_{(2)}\z \mbox{(summation
is understood)\/.}\\ (\Delta \otimes i\!d)\Delta(a) \equiv a_{(1)}
\otimes a_{(2)}
\otimes a_{(3)}\z \mbox{\em etc.}
\end{array}
\label{sweedler}
\end{equation}
We call two Hopf algebras \U\ and \A\ dually paired
if there exists a non-degenerate inner product $<\;,\;>:$
$\U \otimes \A \rightarrow k$, such that
\begin{eqnarray}
<x y,a> & = & <x \otimes y, \Delta(a)> \equiv <x,a_{(1)}><y,a_{(2)}>,
\label{multinduced}\\
<x,a b> & = & <\Delta(x),a \otimes b> \equiv <x_{(1)},a><x_{(2)},b>,
\label{inducedmult}\\
<S(x),a> & = & <x,S(a)>,\\
<x,1> & = & \epsilon(x),\z \mbox{and}\z <1,a>=\epsilon(a),
\end{eqnarray}
for all  $x,y \in \U$ and $a,b \in \A$, {\em i.e.} if the product of
the first Hopf algebra induces the coproduct on the second and
vice versa.
Note that a Hopf algebra is in general
non-cocommutative, {\em i.e.} $\tau \circ \Delta \neq \Delta$.

\subsection{Cartan Calculus}

The purpose of this article is to generalize the Cartan calculus of
ordinary {\em commutative} differential geometry to the case of quantum Lie
algebras.  Before restricting ourselves to this, however, we
first consider an arbitrary Hopf algebra, and for this more general case
we will introduce an exterior derivative \dg,
Lie derivatives $\Li_x$ and inner derivations $\I_x$.  The guideline for
our generalization will be the classical Cartan identity
\begin{equation}
\Li_x = \I_x  \dg + \dg  \I_x,  \label{Cartan}
\end{equation}
the Leibniz rule
\begin{equation}
\dg(a b) = \dg(a) b + a \dg(b)\footnote{We use parentheses
to delimit operations like \dg, $\I_x$ and $\Li_x$, {\em e.g.} $\dg a = \dg(a)
+ a \dg$.  However, if the limit of the operation is clear from the context,
we will
suppress the parentheses, {\em e.g.} $\dg(\I_x \dg a) \equiv
\dg(\I_x(\dg(a)))$.},
\label{Leibniz}
\end{equation}
and the nilpotency of \dg.
In the following we will work with a Hopf algebra \U\, which will be
interpreted as an algebra of left-invariant differential operators or
vector fields, and the dually
paired Hopf algebra \A\, which in this interpretation is the algebra
of functions on which elements of \U\ act via
\begin{equation}
x \tr a = a_{(1)} < x, a_{(2)} >,
\end{equation}
where $x \in \U$ and $a \in \A$.
The action of $x$ on a pair of functions $a$, $b$ $\in \A$ is given
in terms of the coproduct by
\begin{equation}
x \tr a b  = (x_{(1)} \tr a) (x_{(2)} \tr b). \label{XAB}
\end{equation}
This motivates the introduction of a product structure on the ``cross
product'' algebra \A\cross\U\ \cite{SW,SWZ3} via the commutation relation
\begin{equation}
x a = a_{(1)} < x_{(1)}, a_{(2)} > x_{(2)}.
\end{equation}
As in the classical case, the Lie derivative of a function is
given by the action of the corresponding vector field, {\em i.e.}
\begin{equation}
\begin{array}{l}
\Li_x(a) = x \tr a =  a_{(1)} < x, a_{(2)} >,\\
\Li_x a = a_{(1)} < x_{(1)}, a_{(2)} > \Li_{x_{(2)}}.
\end{array} \label{XA}
\end{equation}
The Lie derivative along $x$ of an element $y \in  \U$ is given by the
adjoint action in \U:
\begin{equation}
\Li_x(y) = x \ad y =  x_{(1)} y S(x_{(2)}).
\label{xady}
\end{equation}
To find the action of $\I_x$ we can now attempt to use the Cartan
identity (\ref{Cartan})\footnote{The idea is
to use this identity as long as it is consistent
and modify it when needed.}
\begin{equation}
\begin{array}{rcl}
x \tr a & = & \Li_x(a)\\
        & = & \I_x(\dg a) + \dg(\I_x a).
\end{array}
\end{equation}
As the inner derivation $\I_x$ contracts 1-forms and is zero on
0-forms like $a$, we find
\begin{equation}
\I_x(\dg a) = x \tr a = a_{(1)} < x, a_{(2)} >. \label{incomplete}
\end{equation}
However, this cannot be true for any $x \in \U$ because from the Leibniz
rule for \dg\ we
have $\dg(1) = \dg(1\cdot  1) = \dg(1) 1 + 1 \dg(1) = 2 \dg(1)$
and any $\I_x$ that gives a non-zero result
upon contracting $\dg(1)$ will hence lead to a contradiction. From
(\ref{incomplete}) we see that the troublemakers are $x \in \U$
with $\epsilon(x) \neq 0$. Noting that $\epsilon(x -1 \epsilon(x)) = 0$,
we modify equation (\ref{incomplete}) to read
\begin{equation}
\I_x(\dg a) = a_{(1)} < x -1 \epsilon(x), a_{(2)} >,
\label{ida}
\end{equation}
such that $\I_x(\dg 1) = 1 < x -1 \epsilon(x), 1> = 0$.
Without loss of generality we can now set
\begin{equation}
\dg(1) \equiv 0\x\mbox{and}\x \I_1 \equiv  0.
\end{equation}
For $x \in \U$ with non-zero counit we also need to modify equation
(\ref{Cartan}) to
\begin{equation}
\Li_{x - 1 \epsilon(x)} = \I_x  \dg + \dg  \I_x,
\end{equation}
or in view of (\ref{XA}) identifying $\Li_1 \equiv 1$ and using the
linearity of the Lie derivative,
\begin{equation}
\Li_x = \I_x  \dg + 1 \epsilon(x) + \dg  \I_x\z
\mbox{\em(generalized Cartan identity)}.
\label{gCi}
\end{equation}
Next consider for any form $\al$
\begin{equation}
\begin{array}{rcl}
\Li_x(\dg \al ) & = & \dg(\I_x \dg \al ) + \epsilon(x) \dg(\al )
                      + \I_x(\dg \dg \al )\\
                & = & \dg(\Li_x \al ) + 0,
\end{array}
\label{Ld}
\end{equation}
which shows that Lie derivatives commute with the exterior
derivative:
\begin{equation}
\Li_x  \dg = \dg  \Li_x.
\end{equation}
{}From this and (\ref{XA}) we find
\begin{equation}
\Li_x \dg(a) = \dg(a_{(1)}) < x_{(1)}, a_{(2)} > \Li_{x_{(2)}}.
\end{equation}
To find the complete commutation relations of $\I_x$ with functions
and forms rather than just its action on them
we next compute the action of $\Li_x$ on a product of functions
$a$, $b$ $\in \A$
\begin{equation}
\begin{array}{rcl}
\Li_x(a b) & = & \I_x \dg(a b) + \epsilon(x) a b\\
           & = & \I_x(\dg(a) b + a \dg(b)) + \epsilon(x) a b
\end{array}
\end{equation}
and compare with equation (\ref{XAB}).
In place of an arbitrary $x \in \U$ let us for the moment
specialize to the case of a set of left-invariant
vector fields $\chi_i \in \U$, $i=1\ldots n$, with zero counit
and with coproduct
\begin{equation}
\Delta\chi_i = \chi_i \otimes 1 + O_i{}^j \otimes \chi_j;\z O_i{}^j \in
\U.\label{DO}
\end{equation}
A coproduct of this form is encountered in many known examples and,
as we shall later see, it is not hard to generalize the equations.  For
this choice of vector fields we obtain
\begin{equation}
\begin{array}{rcl}
\I_{\chi_i} a &=& (O_i{}^j \tr a) \;\I_{\chi_j}\\
&=& \Li_{O_i{}^j}(a) \;\I_{\chi_j},
\end{array}
\end{equation}
if we assume that the commutation relation of $\I_{\chi_i}$ with $\dg(a)$ is
of the general form
\begin{equation}
\I_{\chi_i} \dg(a) = \underbrace{\I_{\chi_i}(\dg a)}_{\in \A} +
\mbox{``braiding term''}\cdot\I_{\chi_?}\,.
\end{equation}
A calculation of $\Li_{\chi_i}(\dg(a) \dg(b))$ along similar lines
gives in fact
\begin{equation}
\begin{array}{rcl}
\I_{\chi_i} \dg(a) &=& (\chi_i \tr a) - \dg(O_i{}^j \tr a)
\;\I_{\chi_j}\\
&=&\I_{\chi_i}(\dg a) - \Li_{O_i{}^j}(\dg a) \;\I_{\chi_j}
\end{array}
\end{equation}
and we propose for any $p$-form $\al$:
\begin{equation}
\I_{\chi_i} \al = \I_{\chi_i}(\al) +  (-1)^p \Li_{O_i{}^j}(\al)\;
\I_{\chi_j}.
\end{equation}
Now we are ready to generalize to an arbitrary $x \in \U$ instead of
$\chi_i$. We observe that $\I_x = \I_{(x - 1 \epsilon(x))}$ because of
$\I_1 = 0$ and the linearity of the inner derivation. The special
coproduct given in (\ref{DO}) can now be replaced by
\begin{equation}
\begin{array}{rcl}
\Delta(x - 1 \epsilon(x)) &=& x_{(1)} \otimes x_{(2)} - 1 \epsilon(x)
\otimes 1\\
&=& (x - 1 \epsilon(x)) \otimes 1 + x_{(1)} \otimes (x_{(2)} - 1
\epsilon(x_{(2)}))
\end{array}
\end{equation}
leading to
\begin{equation}
\I_{(x - 1 \epsilon(x))} \al = \I_{(x - 1 \epsilon(x))}(\al) +
(-1)^p \Li_{x_{(1)}}(\al)\;\I_{(x_{(2)} - 1 \epsilon(x_{(2)}))},
\end{equation}
which is equivalent to
\begin{equation}
\I_x \al = \I_x(\al) +(-1)^p \Li_{x_{(1)}}(\al)\;\I_{x_{(2)}}.
\label{IXA}
\end{equation}
For a more direct argument we could also use the requirement
that (\ref{XA}) and (\ref{gCi}) be mutually consistent:  the left-hand
side of the second equation in (\ref{XA}) gives (using the Leibniz rule)
\begin{equation}
\Li_x a =\I_x \dg (a)+\I_x a \dg +\epsilon (x)a +\dg \I_x a
\end{equation}
and the right-hand side gives
\begin{eqnarray}
\lefteqn{a_{(1)}<x_{(1)},a_{(2)}>\Li_{x_{(2)}} =} \nonumber \\
& &a_{(1)}<x_{(1)},a_{(2)}>\dg \I_{x_{(2)}} \\
& &+a_{(1)}<x,a_{(2)}>+a_{(1)}<x_{(1)},a_{(2)}>\I_{x_{(2)}} \dg.\nonumber
\end{eqnarray}
Equating the two and using (\ref{XA}), (\ref{ida}), (\ref{Ld}), and
$\I_x (a)=0$, we find
\begin{equation}
\I_x \dg (a)-\I_x (\dg a)+\Li_{x_{(1)}} (\dg a)\;\I_{x_{(2)}}=-{[\I_x a
-\I_x (a)-\Li_{x_{(1)}} (a)\; \I_{x_{(2)}},\dg]}_+.
\end{equation}
Therefore, we propose equation (\ref{IXA})
for any $p$-form $\al$, so that both sides of the above relation
vanish.

Missing in our list are commutation relations of Lie derivatives with
vector fields and inner derivations.
It was shown in \cite{SWZ3} that the product in \U\ can be expressed
in terms of a right coaction $\DA: \U \rightarrow \U \otimes \A$,
denoted $\DA(y) = y^{(1)} \otimes y^{(2)'}$, such that
$x y = y^{(1)} <x_{(1)},y^{(2)'}> x_{(2)}$.
In the context of (\ref{xady}), this gives
\begin{eqnarray}
\Li_x(y) &=& x_{(1)} y S(x_{(2)}) = y^{(1)} <x,y^{(2)'}>,\\
\Li_x \Li_y &=& \Li_{\mbox{\small \pounds}_{x_{(1)}}(y)} \Li_{x_{(2)}} =
\Li_{y^{(1)}} <x_{(1)},y^{(2)'}> \Li_{x_{(2)}},
\end{eqnarray}
and --- using the Cartan identity ---
\begin{equation}
\Li_x \I_y  = \I_{\mbox{\small \pounds}_{x_{(1)}}(y)} \Li_{x_{(2)}} =
\I_{y^{(1)}} <x_{(1)},y^{(2)'}> \Li_{x_{(2)}}.
\end{equation}

Here is a summary of commutation relations valid on any form:
$x,y \in \U$, $a \in \A$,
$\al$ is a $p$-form and $v \in \A\cross\U$ is a vector field.
\begin{eqnarray}
\Li_x a & = & a_{(1)} < x_{(1)}, a_{(2)} > \Li_{x_{(2)}}\\
\Li_x \dg(a) & = & \dg(a_{(1)}) < x_{(1)}, a_{(2)} > \Li_{x_{(2)}}\\
\Li_x \al & = & \Li_{x_{(1)}}(\al)\:\Li_{x_{(2)}}\\
\I_x a & = & a_{(1)} <x_{(1)},a_{(2)}> \I_{x_{(2)}}\\
\I_x \dg(a) & = & a_{(1)} < x -1 \epsilon(x), a_{(2)} > -
\dg(a_{(1)}) <x_{(1)},a_{(2)}> \I_{x_{(2)}}\\
\I_x \al & = & \I_x(\al) +(-1)^p
\Li_{x_{(1)}}(\al)\;\I_{x_{(2)}}\\
\dg \al & = & \dg(\al) + (-1)^p \al \dg\\
\dg \dg(\al) & = & - (-1)^p \dg(\al) \dg\\
\nonumber \\
\Li_x(v) & = & x_{(1)} v S(x_{(2)}) \label{lixv}\\
\nonumber \\
\dg^2&=&0\\
\dg \Li_x &=& \Li_x \dg\\
\Li_x &=& \dg \I_x + 1 \epsilon(x) + \I_x \dg\\
\Li_x \Li_y &=& \Li_{y^{(1)}} <x_{(1)},y^{(2)'}> \Li_{x_{(2)}}\\
\Li_x \I_y &=& \I_{y^{(1)}} <x_{(1)},y^{(2)'}> \Li_{x_{(2)}}
\end{eqnarray}

\subsection{Maurer-Cartan Forms}

The most general left-invariant 1-form can be written \cite{W2}
\begin{equation}
\om_b := S(b_{(1)}) \dg(b_{(2)}) = - \dg(S b_{(1)} ) b_{(2)}
\end{equation}
\begin{equation}
(\mbox{\em left-invariance:}\x\AD(\om_b) =
S(b_{(2)}) b_{(3)} \otimes S(b_{(1)}) \dg(b_{(4)})
= 1 \otimes \om_b ),
\end{equation}
corresponding to a function $b \in \A$. If this function happens to
be $t^i{}_k$, where $t \in M_n(\A)$ is an $n \times n$ matrix
representation of \U\
with $\Delta (t^i{}_k) =$\mbox{$t^i{}_j \otimes t^j{}_k$} and
$S(t)=t^{-1}$, we obtain
the well-known Cartan-Maurer form $\om_t = t^{-1} \dg(t)$. Here is a
nice formula for the exterior derivative of $\om_b$:
\begin{equation}
\begin{array}{rcl}
\dg(\om_b) & = & \dg(S b_{(1)} ) \dg(b_{(2)})\\
&=& \dg(S b_{(1)} ) b_{(2)} S(b_{(3)}) \dg(b_{(4)})\\
&=& - \om_{b_{(1)}} \om_{b_{(2)}}.
\end{array}
\end{equation}
The Lie derivative is
\begin{equation}
\begin{array}{rcl}
\Li_x(\om_b)&=&\Li_{x_{(1)}}(S b_{(1)} ) \Li_{x_{(2)}}(\dg b_{(2)})\\
&=&<x_{(1)},S(b_{(1)})> S(b_{(2)}) \dg(b_{(3)}) <x_{(2)},b_{(4)}>\\
&=&\om_{b_{(2)}} <x,S(b_{(1)})b_{(3)}>\\
&=&<x_{(1)},S(b_{(1)})> \om_{b_{(2)}} <x_{(2)},b_{(3)}>.
\end{array} \label{XOM}
\end{equation}
The contraction of left-invariant forms with $\I_x$
--- {\em i.e.} by a {\em left-invariant} $x \in \U$ ---
gives a number in the field $k$ rather than a function
in \A\ as was the case for $\dg(a)$:
\begin{equation}
\begin{array}{rcl}
\I_x(\om_b) & = & \I_x(- \dg(S b_{(1)} ) b_{(2)})\\
&=& - \I_x(\dg S b_{(1)} ) b_{(2)}\\
& = & -< x -1 \epsilon(x),S(b_{(1)})> S(b_{(2)}) b_{(3)}\\
&=& -<x,S(b)> + \epsilon(x) \epsilon(b).
\end{array} \label{IOM}
\end{equation}
As an exercise and to check consistency we will compute the
same expression in a different way:
\begin{equation}
\begin{array}{rcl}
\I_x(\om_b) & = & \I_x(S(b_{(1)}) \dg(b_{(2)}))\\
& =& <x_{(1)},S(b_{(1)})> S(b_{(2)}) \I_{x_{(2)}}(\dg b_{(2)})\\
& = & <x_{(1)},S(b_{(1)})> S(b_{(2)}) b_{(3)} <x_{(2)} - 1
\epsilon(x_{(2)}),b_{(4)}>\\
& =& <x_{(1)},S(b_{(1)})><x_{(2)} - 1 \epsilon(x_{(2)}),b_{(2)}>\\
&=&  \epsilon(x) \epsilon(b) -<x,S(b)>.
\end{array}
\end{equation}

\section{Quantum Lie Algebras}

We now turn our attention to the case where the Hopf algebras in
question are quantum Lie algebras and the functions on the quantum group
corresponding to that algebra.  We start with a Lie algebra {\bf g};
its {\em quantum universal
enveloping algebra} \uqg\ is the Hopf algebra whose elements are
polynomials in the generators of {\bf g} modulo (deformed) commutation
relations.
Dually paired with \uqg\ is \fun, the Hopf algebra of
functions on the quantum group {\bf G}$_q$.

In the following we would like to concentrate on a bicovariant basis
of left-invariant vector fields \cite{SWZ3}
 $\{ \chi_i \in \uqg| i=1,\ldots ,n \}$, {\em i.e.} the $\chi_i$s are
left-invariant
and close under right coaction:
\begin{eqnarray}
\AD(\chi_i) &=& 1\otimes \chi_i, \nonumber \\
\DA(\chi_i) &=& \chi_j \otimes T^j{}_i,
\end{eqnarray}
with $T \in M_n(\A)$.
The identification with {\bf g} is made by requiring that, in the
classical limit, the $\chi_i$s reduce to either a generator of
{\bf g} or a Casimir operator in $U(\mbox{\bf g})$.
(Note that such a choice of basis might not be unique or even possible.)
We interpret ${\cal T}_q :=$span$\{\chi_i\}$ as the quantum analogue of a
tangent bundle, and
take \U\ (see previous section) to be its UEA.

Dual to this basis of vector fields is a basis
of 1-forms $\om^i \equiv \om_{b^i}$ corresponding to a set of functions
$b^i \in \A$ that satisfy
\begin{equation}
\I_{\chi_i}(\om^j) = -<\chi_i,S(b^j)> = \delta_i^j.
\end{equation}
${\cal T}^*_q :=$span$\{\om^j\}$ is to be interpreted as the quantum
analogue of the cotangent bundle.
Using these particular 1-forms we can reexpress the exterior derivative
on functions $a \in \A$ as
\begin{equation}
\dg(a) = \om^i (\chi_i \tr a) = \om^i \Li_{\chi_i}(a).
\end{equation}
{}From the previous equation and the Leibniz rule for $\dg$, we
find
\begin{equation}
a \om^i = \om^j \Li_{\Theta_j{}^i}(a),
\end{equation}
where
\begin{equation}
\Theta_j{}^i = -{\chi_j}_{(1)} < {\chi_j}_{(2)} , S(b^i) > + \chi_j
\epsilon(b^i).
\end{equation}
Furthermore, the requirement that \dg\ be invariant under coactions
gives an (often overlooked) additional condition on the
$b^i$'s:
\begin{equation}
\DA(\om^i) = \om^j \otimes S^{-1}(T^i{}_j),
\end{equation}
and therefore
\begin{equation}
\Delta^{\mbox{\small Ad}}(b^i) = b^j \otimes S^{-1}(T^i{}_j),
\end{equation}
where $\Delta^{\mbox{\small Ad}}(b) \equiv b_{(2)} \otimes S(b_{(1)})
b_{(3)}$, and we have used $\DA(\om_b) = \om_{b_{(2)}} \otimes
S(b_{(1)}) b_{(3)}$.
If we assume for simplicity a coproduct of the standard form
\begin{equation}
\Delta (\chi_i) = \chi_i \otimes 1 + O_i{}^j \otimes \chi_j,
\label{standard}
\end{equation}
where $O_i{}^j \in \uqg$, then
from (\ref{XOM}) and (\ref{IOM}) we find commutation relations
for $\I_{\chi_i}$ with $\om^j$,
\begin{equation}
\begin{array}{rcl}
\I_{\chi_i} \om^j &=& \delta_i^j - \Li_{O_i{}^k}(\om^j) \I_{\chi_k}\\
&=& \delta_i^j - \om^m <O_i{}^k,S^{-1}(T^j{}_m)> \I_{\chi_k},
\end{array} \label{IOMI}
\end{equation}
which can be used to define the wedge product $\wedge$ as some kind of
antisymmetrized tensor product\footnote{So far we have
suppressed the $\wedge$-symbol; to avoid confusion we will reinsert it
in this paragraph.}:  as in the classical case we make an ansatz for
the product of two forms in terms of tensor products
\begin{equation}
\om^i \wedge \om^j = \om^i \otimes \om^j - \hat{\sigma}^{ij}{}_{mn}
\om^m \otimes \om^n,
\end{equation}
with as yet unknown numerical constants $\hat{\sigma}^{ij}{}_{mn} \in
k$, and define $\I_{\chi_i}$ to act on this product by contracting
in the first tensor product space, {\em i.e.}
\begin{equation}
\I_{\chi_i}(\om^j \wedge \om^k) = \delta_i^j \om^k -
\hat{\sigma}^{jk}{}_{mn} \delta_i^m \om^n.
\end{equation}
But from (\ref{IOMI}) we already know how to compute this,
namely
\begin{equation}
\begin{array}{rcl}
\I_{\chi_i}(\om^j \wedge \om^k)&=&\delta_i^j \om^k -
\Li_{O_i{}^{\ell}}(\om^j) \delta_{\ell}^k\\
&=& \delta_i^j \om^k - \om^m <O_i{}^k,S^{-1}(T^j{}_m)>
\end{array}
\end{equation}
and by comparison we find
\begin{equation}
\hat{\sigma}^{ij}{}_{mn} = <O_m{}^j,S^{-1}(T^i{}_n)>,
\end{equation}
or
\begin{equation}
\begin{array}{rcl}
\om^i \wedge \om^j &=& \om^i \otimes \om^j - <O_m{}^j,S^{-1}(T^i{}_n)>
\om^m \otimes \om^n\\
&=& (I-\hat{\sigma})^{ij}{}_{mn}\om^m \otimes \om^n \\
&=& \om^i \otimes \om^j - \om^k \otimes \Li_{O_k{}^j}(\om^i).
\end{array}
\end{equation}
These equations can be used to obtain the (anti)commutation relations
between the $\om^i$s; by using the characteristic equation for
$\hat{\sigma}$, projection matrices orthogonal to the antisymmetrizer $I-
\hat{\sigma}$ can be found, and these will annihilate $\om^i \wedge
\om^j$.  The resulting equations will determine how to commute the
1-forms.
Using the same method we can also obtain a tensor product decomposition
of products of inner derivations
\begin{equation}
\I_m \wedge \I_n = \I_m \otimes \I_n - \hat{\sigma}^{ij}{}_{mn}
\I_i \otimes \I_j,
\end{equation}
defined to act on 1-forms by contraction in the first tensor product
space.
This can again be used to find (anti)commutation relations for the
$\I$s via projection matrices as mentioned above.
\\{\em Remark:} The tensor product decomposition of the
wedge product is invariant
under linear changes of the $\{\chi_i\}$ basis, but it does
depend on our choice of quantum tangent bundle.

There is actually an operator $W$  that recursively translates
wedge products into the tensor product representation:
\begin{equation}
\begin{array}{l}
W: \Lambda^p_q \rightarrow
{\cal T}^*_q \otimes \Lambda^{p-1}_q,\x p \geq 1,\\
W(\alpha) = \om^n \otimes \I_{\chi_n}(\alpha)
\end{array}
\end{equation}
for any p-form $\alpha$, {\em e.g.}
\begin{equation}
\begin{array}{rcl}
\om^n \otimes \I_{\chi_n}(\om^j \wedge \om^k) & = & \om^n \otimes
(\delta^j_n \om^k - \Li_{O_n{}^m}(\om^j) \delta^k_m)\\
&  = & \om^j \otimes \om^k - \om^n \otimes \Li_{O_n{}^k}(\om^j).
\end{array}
\end{equation}

$W$ is {\em not} limited to Hopf algebras with a coproduct of the
standard form (\ref{standard}); any form of the coproduct is admissible.
For example, the matrix $\hat{\sigma}$ will generalize to
\begin{equation}
\hat{\sigma}^{ij}_{mn} = - <\chi_m,S^{-1}(T^i{}_n)S(b^j)>
                         + \epsilon(b^j) <\chi_m,S^{-1}(T^i{}_n)>.
\end{equation}
This is a generalization of results in \cite{W2}. However, examples
of bicovariant vector fields with comultiplication more general
than (\ref{standard}) have not been studied yet.

\section{Conclusion}

In this article we were able to define the actions of an exterior
derivative, Lie derivatives and inner derivations on forms and
non-commutative functions such that these objects satisfy a closed
algebra, namely a generalized Cartan Calculus.  Most of the relations
that we derive require only a Hopf algebra structure. To be able to
give commutation relations for inner derivations, forms, and forms
with functions, however, we need to make reference to a (finite)
bicovariant basis of left-invariant vector fields. Such a bicovariant
basis permits the decomposition of wedge products into tensor products
as well as a realization of the exterior derivative in terms of
1-forms and vector fields. It is interesting to observe that all
the ``braiding'' was done by Lie derivatives, {\em e.g.} $\Li_{O_i{}^j}$.

We focused on Lie derivatives and inner derivations along
{\em left-invariant} vector fields, {\em i.e.}
elements of \U. This approach is both a generalization and a
restriction of the undeformed theory: the classical case only involves
derivatives along vector fields in the tangent bundle ({\em e.g.}
$X \in$ span({\bf g})) but allows functional coefficients, {\em i.e.} the
vector fields need not be left-invariant.  In
contrast to this we introduce derivatives along any element
of the UEA. Noting that $1 + X$ and even $e^X$ are such
elements, the name ``transport'' instead of ``derivative'' might be
more appropriate.  An attempt to introduce derivatives along
elements in $\A\cross\U$ leads us to the following set of equations:
\begin{eqnarray}
\I_{f \chi} & = & f \I_{\chi}\\
\Li_{f \chi} & = & \dg f \I_\chi +  f \I_\chi \dg + \epsilon(\chi)
\Li_{f}\\
\Li_{f \chi} & = & f \Li_{\chi} + \dg(f) \I_\chi +
\epsilon(\chi) (\Li_f - f) \label{liefx}\\
\Li_{f \chi} \dg & = & \dg \Li_{f \chi}
\end{eqnarray}
The range of validity of these equations is rather
limited; if, for instance, we allow $\chi$ to have non-zero
counit then the aforementioned formulas seem to be only consistent
when evaluated on $a$ or $\dg(a)$, where $a \in \A$ is an arbitrary
function. If, however, we consider only $\chi$s with zero counit
then the problematic term $\Li_{f}$ drops out of all
equations. Equation (\ref{liefx}) becomes
\begin{equation}
\Li_{f \chi}  =  f \Li_{\chi} + \dg(f) \I_\chi, \z \epsilon(\chi) = 0,
\end{equation}
and can be used to define Lie derivatives recursively on any form.
There does not seem to be a way to generalize (\ref{lixv}), {\em i.e.} to
introduce Lie derivatives of {\em vector fields} along {\em
arbitrary} elements of $\A\cross\U$ in the quantum case. Exceptions
will be discussed in \cite{CS}.

\newpage

%change the bibliography as needed


\begin{thebibliography}{199}
%\bibitem{A} E. Abe, \underline{Hopf Algebras} (Cambridge Univ. Press,
%1980)
%\bibitem{Mn1} Yu. Manin, Commun. Math. Phys. {\bf 123} 163 (1989)
%\bibitem{WZ} J. Wess and B. Zumino, Nucl. Phys. B (Proc. Suppl.) {\bf 18B} 302
%(1990)
%\bibitem{Z1} B. Zumino, Mod. Phys. Lett. A {\bf 13} 1225 (1991)
%\bibitem{Z2} B. Zumino, (Berkeley preprint, LBL-31432 and UCB-PTH-59/91),
%in: K.Schm\"udgen (Ed.), Math. Phys. X,
%Proc. X-th IAMP Conf. Leipzig (1991),
%Springer-Verlag (1992)
%\bibitem{Mn2} Yu. Manin, Bonn preprint MPI/91-47
%\bibitem{Mn3} Yu. Manin, Bonn preprint MPI/91-60
%\bibitem{Ms1} G. Maltsiniotis, C. R. Acad. Sci. Paris, {\bf 331} 831 (1990)
%\bibitem{Ms2} G. Maltsiniotis, ``Calcul differentiel sur le groupe
%lin\'{e}aire quantique'', ENS expos\'{e} (1990)
%\bibitem{S} A. Schirrmacher, Munich preprint, MPI-PTH-91-117, presented at the
%1st Max Born Symp. in Theoretical Physics, Wroclaw, Poland, Sept. 27-29, 1991
%\bibitem{S1} A. Sudbery, York preprint PRINT-91-0498 (YORK)
%\bibitem{S2} A. Sudbery, York preprint YORK-92-1
%\bibitem{CW} U. Carow-Watamura, M. Schlieker, S. Watamura, and W. Weich,
%Commun.Math. Phys. {\bf 142} 605 (1991)
%\bibitem{W2} S. L. Woronowicz, Commun. Math. Phys. {\bf 122} 125 (1989)
%\bibitem{Md1} S. Majid, Int. J. Mod. Phys. {\bf A 5} 1 (1990)
%\bibitem{Sw} M. E. Sweedler, \underline{Hopf Algebras} (Benjamin, 1969)
%\bibitem{Df} V. G.Drinfel'd, Proc. Int. Congr. Math., Berkeley 798 (1986)
%\bibitem{W1} S. L. Woronowicz, Commun. Math. Phys. {\bf 111} 613 (1987)
%\bibitem{RTF} N. Yu. Reshetikhin, L. A. Takhtadzhyan, and L. D. Faddeev,
%Leningrad Math. J. {\bf 1} 193 (1990)
%\bibitem{SWZ} P. Schupp, P.Watts, and B. Zumino, Lett. Math. Phys. {\bf 25}
%139 (1992)
%\bibitem{Rn} N. Yu. Reshetikhin, Leningrad Math. J. {\bf 1} (2) 491 (1990)
%\bibitem{RSTS} N. Yu. Reshetikhin, M. A. Semenov-Tian-Shansky, Lett. Math.
%Phys. {\bf 19} (1990)
%\bibitem{J} B. Jur\v{c}o, Lett. Math. Phys. {\bf 22} 177 (1991)
%\bibitem{Ku} P. P. Kulish, R. Sasaki, C. Schwiebert, Preprint YITP/U-92-07
%(March 1992)
%\bibitem{Md2} S. Majid, Preprint, DAMTP/92-12
%\bibitem{Z2} B. Zumino, in: K.Schm\"udgen (Ed.), Math. Phys. X,
%Proc. X-th IAMP Conf. Leipzig (1991),
%Springer-Verlag (1992)
%\bibitem{C1} C. Chryssomalakos, Mod. Phys. Lett. {\bf A}, and
%Preprint, LBL-32442
%\bibitem{C2} C. Chryssomalakos, B. Drabant, M. Schlieker, W. Weich, B. Zumino,
%Commun. Math. Phys. {\bf 147} (3) 635 (1992)
%\bibitem{DJS} B. Drabant, B. Jur\v{c}o, M. Schlieker, W. Weich, B. Zumino,
%Preprint UCB-PTh 92/16 and LBL-32354
%\bibitem{Md3} S. Majid, Preprint, DAMTP/92-10
%\bibitem{B} D. Bernard, Prog. of Theoretical Phys. Supp. {\bf 102} 49 (1990)
%\bibitem{CC} P. Aschieri, L. Castellani, Preprint, CERN-TH.6565/92
%\bibitem{W3} S.L. Woronowicz, Lett. Math. Phys. {\bf 23} 251 (1991)
%\bibitem{SWZ2} P. Schupp, P. Watts, B. Zumino, Lett. Math. Phys {\bf 24} 141
%(1992)
%\bibitem{SW} P.Schupp, P. Watts, preprint, LBL-33274 and
%UCB-PTH-92/42, submitted to Lett. Math. Phys.
%\bibitem{SWZ3} P. Schupp, P. Watts, B. Zumino, preprint, LBL-32315 and
%UCB-PTH-92/14, to be published in  Commun. Math. Phys.
%%%%%%%%%%%%%%%%%%%%%%%%%%%%%%%%%%%%%%%%%%%%%%%%%%%%%%%%%%%%%%%%%%%%%%%%
\bibitem{W2} S. L. Woronowicz, Commun. Math. Phys. {\bf 122} 125 (1989)
\bibitem{RTF} N. Yu. Reshetikhin, L. A. Takhtadzhyan, L. D.
Faddeev, Leningrad Math. J. {\bf 1} 193 (1990)
\bibitem{SWZ} P. Schupp, P. Watts, B. Zumino, Lett. Math. Phys. {\bf
25} 139 (1992)
\bibitem{B} D. Bernard, Prog. of Theoretical Phys. Supp. {\bf 102} 49
(1990)
\bibitem{CC} P. Aschieri, L. Castellani, preprint, CERN-TH.6565/92
\bibitem{CW} U. Carow-Watamura, M. Schlieker, S. Watamura, W. Weich,
Commun. Math. Phys. {\bf 142} 605 (1991)
\bibitem{Md1} S. Majid, Int. J. Mod. Phys. {\bf A 5} 1 (1990)
\bibitem{SW} P. Schupp, P. Watts, preprint, LBL-33274 and
UCB-PTH-92/42, submitted to Lett. Math. Phys.
\bibitem{SWZ3} P. Schupp, P. Watts, B. Zumino, preprint, LBL-32315 and
UCB-PTH-92/14, to be published in  Commun. Math. Phys.
\bibitem{A} E. Abe, \underline{Hopf Algebras} (Cambridge Univ. Press,
1977)
\bibitem{Sw} M. E. Sweedler, \underline{Hopf Algebras} (Benjamin, 1969)
\bibitem{Df} V. G. Drinfel'd, Proc. Int. Congr. Math., Berkeley 798
(1986)
\bibitem{CS} C. Chryssomalakos, P. Schupp, ``Vector Fields on
Quantum Groups'' (in preparation)
\end{thebibliography}
\end{document}